\pdfoutput=1
\documentclass[a4paper,aps,prstab,preprint,superscriptaddress]{revtex4-1}
\usepackage[utf8]{inputenc}
\usepackage{amsmath}
\usepackage{amsfonts}
\usepackage{amssymb}
\usepackage{graphicx}
\usepackage{natbib}

\def\a{{\mathrm a}}
\def\b{{\mathrm b}}

\def\d{{\mathrm d}}

\def\g{{\mathrm g}}

\def\i{{\mathrm i}}

\def\m{{\mathrm m}}

\def\p{{\mathrm p}}

\def\r{{\mathrm r}}

\def\x{{\mathrm x}}

\newcommand{\mi}{\mathrm{i}} 

\begin{document}

\author{C. Arran}
\affiliation{John Adams Institute for Accelerator Science, University of Oxford, Denys Wilkinson Building, Keble Road, Oxford OX1 3RH, United Kingdom}
\author{N.H. Matlis}
\affiliation{Deutsches Elektronen-Synchrotron (DESY), Notkestraße 85, Hamburg 22607, Germany}
\author{R. Walczak}
\affiliation{John Adams Institute for Accelerator Science, University of Oxford, Denys Wilkinson Building, Keble Road, Oxford OX1 3RH, United Kingdom}
\author{S.M. Hooker}
\affiliation{John Adams Institute for Accelerator Science, University of Oxford, Denys Wilkinson Building, Keble Road, Oxford OX1 3RH, United Kingdom}

\title{Reconstructing non-linear plasma wakefields using a generalized Temporally Encoded Spectral Shifting analysis}

\begin{abstract}
We generalize the Temporally Encoded Spectral Shifting (TESS) analysis for measuring plasma wakefields using spectral interferometry to dissimilar probe pulses of arbitrary spectral profile and to measuring non-linear wakefields. We demonstrate that the Gaussian approximation used up until now results in a substantial mis-calculation of the wakefield amplitude, by a factor of up to two. A method to accurately measure higher amplitude quasi-linear and non-linear wakefields is suggested, using an extension to the TESS procedure, and we place some limits on its accuracy in these regimes. These extensions and improvements to the analysis demonstrate its potential for rapid and accurate on-shot diagnosis of plasma wakefields, even at low plasma densities. 
\end{abstract}

\maketitle

\section{Introduction}

Acceleration of electrons by plasma wakefields has demonstrated much potential, with laser-driven plasma wakefield acceleration (LWFA) \cite{Tajima1979} producing electrons on the GeV scale over interaction lengths of just a few centimetres \cite{Mangles2004, Leemans2006, Leemans2014}, while beam-driven wakefield acceleration (PWFA) \cite{Chen1985, Rosenzweig1988} has used longer, metre scale plasma cells to boost electron energies by 10s of GeV \cite{Blumenfeld2007,Litos2014}. In order to assist the development of this new technology, it is vital that sensitive, non-invasive diagnostics are developed which are capable of characterizing the wakefield structure and the electron beams they produced.

Measurement of the plasma wakefield is commonly made through changes in the refractive index of the plasma, determined by the effect on an optical probe through techniques such as shadowgraphy \cite{Savert2015} and photon acceleration \cite{Wilks1989}. An optical probe passing through the plasma accumulates a phase change $\delta \phi \propto n_{e0} \int \left( \delta n_e /n_{e0} \right) \d \ell$ \cite{Esarey1990}. At high electron densities this phase change is large even when integrating over short interaction lengths; with electron densities of $10^{18}$ cm$^{-3}$ and wake amplitudes of $ \delta n_e /n_{e0} \sim 0.1$ the phase accumulated is of the order of one radian after an interaction length of just $1$ mm. However, as LWFA experiments aim to increase the energy gain, the length of the acceleration stages is increasing. To do this one must increase the dephasing length $L_{\d\p} \propto n_e^{-\frac{3}{2}}$ by moving to lower plasma densities, around $10^{17}$ cm$^{-3}$ \cite{Leemans2009}\cite{Li2018}. Transverse probing techniques such as shadowgraphy, with an interaction length on the scale of the plasma wavelength $\lambda_\p \propto n_e^{-\frac{1}{2}}$, are no longer sufficiently sensitive, as the total phase change varies as $\delta \phi \propto \lambda_\p \delta n_e \propto n_e^{+\frac{1}{2}}$.

Collective Thomson scattering is a useful diagnostic for measuring waves in these low density plasmas (e.g. \cite{Slusher1980}). However, when probing at an angle to the pump beam the geometry must be carefully matched to the very shallow scattering angle, which is dependent on the plasma density. Furthermore, the scattered power, which varies with the plasma wave amplitude as $P_s \propto \left( \delta n_e /n_{e0} \right)^2 n_{e0}^2$, is extremely small for weak plasma waves at low densities, requiring intense probe pulses or sensitive detectors. It is therefore more common in LWFA experiments to maximise the scattering rate by using a co-propagating geometry, where the plasma wave is probed longitudinally (e.g. \cite{LeBlanc1996}). In this situation, it becomes difficult when using ultrashort broadband probe pulses to separate the scattered light from the original probe spectrum, leaving the Stokes and anti-Stokes spectral components poorly resolved and making it impossible in practice to extract information about the plasma wave. Using longer duration probe pulses would overcome this difficulty, but at the cost of reducing the temporal resolution.

Frequency Domain Interferometry (FDI) \cite{Geindre1994,Siders1996} and Holography (FDH) \cite{Matlis2006} are some of the most sensitive techniques for longitudinally probing rapidly evolving density structures with high temporal resolution. In these methods, co-propagating probe and reference pulses pass through an interaction region along the same path but are separated in time, such that the probe accumulates a phase change due to the density structure which the reference pulse does not. The interaction region is imaged onto the entrance slit of a spectrometer, giving spatial resolution of the density structure in one dimension, while in the spectrometer the separation between the probe and reference pulses leads to spectral interference, with a fringe spacing dependent on the delay between the pulses. The interference pattern contains information about the difference in phase shifts --- and hence the local density of the plasma ---  experienced by the probe and reference pulses.

In both FDI and FDH the probe pulse co-propagates with the plasma wakefield, and hence --- assuming that these waves propagate at the same speed --- these measurements can determine the electron density as a function of a local space coordinate fixed in the frame of the plasma wakefield. In FDI the probe pulse is shorter than the plasma wavelength, and hence scanning the delay between the probe and reference pulses allows the temporal behaviour of the plasma wave to be mapped out. In FDH the probe and reference pulses are chirped and stretched, and hence the temporal dependence of the wakefield can be determined, in a single shot, from the spectral phase of the probe. However, the phase reconstruction analysis process in FDH requires several further measurements of the temporal and spectral phases of both the probe and reference pulses in order to distinguish the wakefield information from the intrinsic phase of the pulses. Recently, the technique of Temporally Encoded Spectral Shifting (TESS) \cite{Matlis2016} was developed to extract information about phase modulation, such as from plasma wakefields, without phase reconstruction. The method uses the same experimental set up as FDH but involves less computationally expensive analysis and requires fewer reference measurements.

To date, however, the TESS analysis has been restricted to the case of probe and reference pulses which have identical Gaussian spectral profiles. Here we extend the TESS method to the case of non-identical probe and reference pulses of arbitrary spectral profile. We use experimental results to demonstrate that under real conditions the assumption of Gaussian probe and reference pulses can lead to significant errors in the deduced wakefield. We also extend the analysis to the case of non-linear plasma wakefields and use simulations to show that the wakefield amplitude and frequency can accurately be recovered for quasi-linear wakefields, and that the wakefield frequency can still be recovered for strongly non-linear wakefields.

\section{Arbitrary Probe and Reference Pulses} \label{sec: General Probe Pulses}

\subsection{TESS in General}

As described by Matlis et al. \cite{Matlis2016}, a probe pulse co-propagating with a linear plasma wave is transformed in a well defined way. In the linear regime the plasma wakefield is sinusoidal, and information about the wakefield is encoded in the modulation of the spectrum of the probe by an additional phase $\phi_\mathrm{wake}(\zeta) = \phi_{1} \sin (\omega_{\p0} \zeta)$, where $\omega_{\p0}$ is the non-relativistic plasma frequency, $\zeta = \tau- z/v_\g$ is a co-moving time coordinate for the probe pulse, which propagates at $v_\g$, and the amplitude of the phase is proportional to the density amplitude of the plasma wave, $\phi_1 = \left({\omega_{\p0}^{~2} L}/{2 \omega_0 c}\right) \left({\delta n_e}/{n_{e0}} \right)$. This leads to the transformations between the electric field of the probe before entering the plasma, in temporal space $E_\mathrm{pr}(\zeta)$ and in spectral space $\mathcal{E}_\mathrm{pr}(\omega)$, and after leaving the plasma, $E'_\mathrm{pr}(\zeta)$ and $\mathcal{E}'_\mathrm{pr}(\omega)$, shown below:

\begin{align}
E_\mathrm{pr}'(\zeta) &= E_\mathrm{pr}(\zeta) e^{\mi \phi_{1} \sin (\omega_{\p0} \zeta} \nonumber \\
&= E_\mathrm{pr} (\zeta) \sum^\infty_{k=-\infty} J_k(\phi_1) e^{\mi k\omega_{\p0} \zeta} . \\
\mathcal{E}_\mathrm{pr}'(\omega) &= \frac{1}{\sqrt{2\pi}} \int^{\infty}_{-\infty} E_\mathrm{pr}(\zeta) e^{\mi \phi_\mathrm{wake}(\zeta)} e^{-\mi \omega \zeta} ~\d \zeta \nonumber \\
&= \sum^\infty_{k=-\infty} J_k(\phi_1) \mathcal{E}_\mathrm{pr} (\omega - k\omega_{\p0}) , \label{eqn: Jacobi-Anger}
\end{align}
where $J_n(x)$ is the $n$-th order Bessel function of the first kind.

\begin{figure*}[h!]
\begin{center}
\includegraphics[scale=0.7]{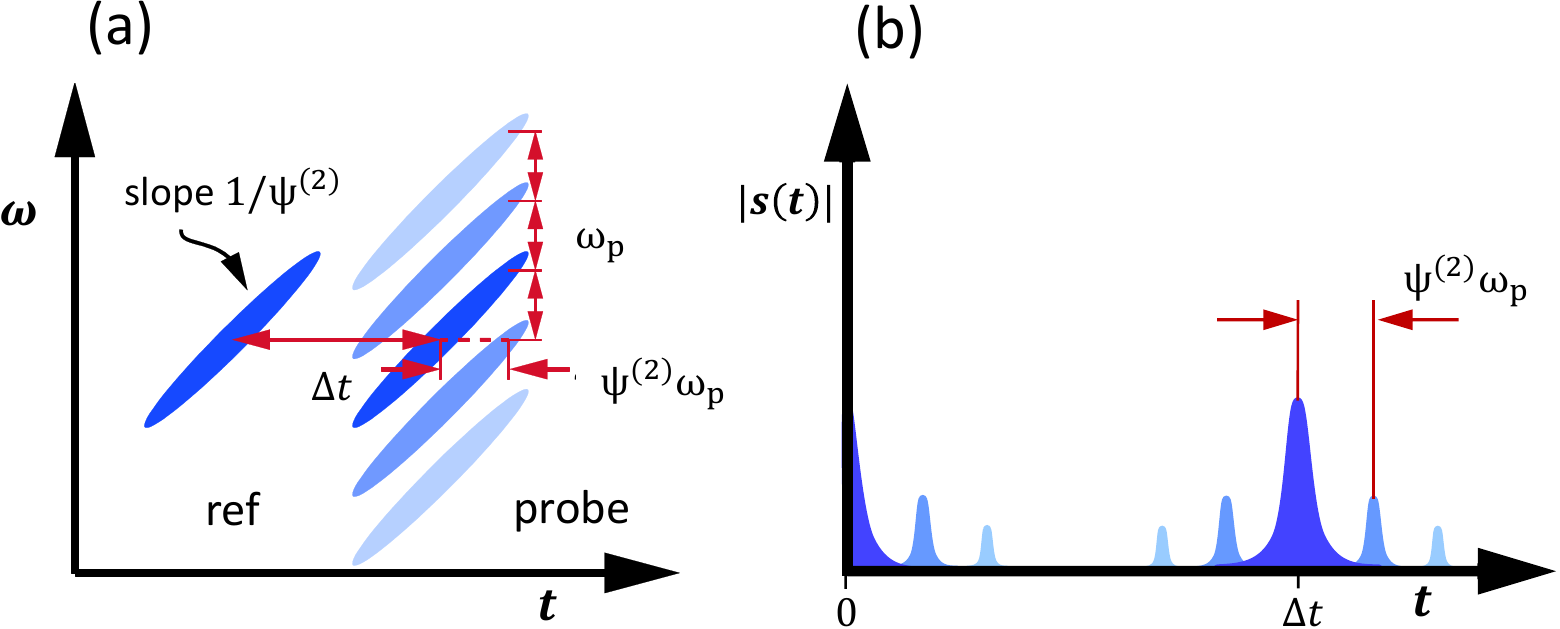}
\caption{a) Frequency-time domain plots of the reference pulse, and the probe pulse after it has interacted with a sinusoidal plasma wave of frequency $\omega_{\p0}$. Modulation of the probe pulse generates copies of the incident probe pulse, spectrally shifted by multiples of $\omega_{\p0}$. b) The TESS signal, obtained by a Fourier transform of the recorded spectrum of the transmitted probe and reference pulses. The temporal separation of the probe and reference pulses yields a DC term at $t=0$ and a sideband at $t = \Delta t$, and modulation of the probe causes a series of satellites (lighter blue) separated from the sideband by multiples of $\omega_{\p0} \psi^{(2)}$}
\label{fig: TESS Schematic}
\end{center}
\end{figure*}

From eqn~\eqref{eqn: Jacobi-Anger} we see that after it interacts with the plasma wave the spectrum of the transmitted probe pulse is a superposition of the original spectrum of the incident pulse and copies shifted in frequency by multiples of the plasma frequency, as shown in the cartoon in Fig.~\ref{fig: TESS Schematic}a). The reference, meanwhile, remains unchanged, co-propagating with the probe a time $\Delta t$ earlier:

\begin{align}
E_\mathrm{r}'(\zeta) &= E_\mathrm{r}(\zeta + \Delta t) . \\
\mathcal{E}_\mathrm{r}'(\omega) &= \frac{1}{\sqrt{2\pi}} \int^{\infty}_{-\infty} E_\mathrm{r}(\zeta+\Delta t) e^{-\mi \omega \zeta} ~\d \zeta \nonumber \\
&= \mathcal{E}_\mathrm{r} (\omega) e^{+\mi \omega \Delta t} .
\end{align}

The two pulses interfere within the spectrometer to create a spectral interferogram $\mathcal{S}(\omega) = \left| \mathcal{E}_\mathrm{pr}'(\omega) \right|^2 + \left| \mathcal{E}_\mathrm{r}'(\omega) \right|^2 + \mathcal{E}_\mathrm{pr}'(\omega) \mathcal{E}_\mathrm{r}'^*(\omega) + c.c.$, where $c.c$. denotes the complex conjugate of the previous term. In FDH the interferogram is analyzed as follows: First, one uses an inverse Fourier transform to the temporal domain to isolate the interference term, $\mathcal{E}_\mathrm{pr}'(\omega) \mathcal{E}_\mathrm{r}'^*(\omega) $, as described in Takeda et al. \cite{Takeda1982}, before Fourier transforming back to the spectral domain. Next, one removes information about the reference pulse to recover the electric field of the transmitted probe, $\mathcal{E}_\mathrm{pr}'(\omega)$. From this, another inverse Fourier transform gives $E'_{\p\r}(\zeta)$; removing the temporal phase of the probe and extracting the phase due to the wakefield one obtains $\phi_\mathrm{wake}(\zeta)$.

In TESS, however, it is only necessary to make the first inverse Fourier transform to the temporal domain. This yields a TESS signal of the form shown schematically in Fig.~\ref{fig: TESS Schematic}b). The form of the signal is given by \cite{Matlis2016}:
\begin{widetext}
\begin{align}
s(t) &= \frac{1}{\sqrt{2\pi}} \int^{\infty}_{-\infty} \mathcal{S}(\omega) e^{+i \omega t}~ \d \omega \nonumber \\
&= H_\mathrm{r,r}(t,0) + \sum_{m=-\infty}^\infty g_m(\phi_1, t, \omega_{\p0}) H_\mathrm{pr,pr}(t,m\omega_{\p0}) \nonumber \\ 
&~~~~ + \sum_{k=-\infty}^\infty J_k(\phi_1) \left[  H_\mathrm{pr,r}(t-\Delta t,k\omega_{\p0}) +  H^*_\mathrm{pr,r}(-t-\Delta t,k\omega_{\p0}) \right] \label{eqn: TESS Signal} \\
\mathrm{where }~~ H_\mathrm{a,b}(t,\Omega) &\equiv \frac{1}{\sqrt{2\pi}} \int^{\infty}_{-\infty}  \mathcal{E}_\mathrm{a}(\omega - \Omega) \mathcal{E}_\mathrm{b}^*(\omega) e^{\mi \omega t}~ \d \omega \label{eqn: H in general} \\
\mathrm{and }~~ g_m(\phi_1,t,\Omega) &\equiv \sum_{n=-\infty}^\infty  J_n(\phi_1) J_{m+n}(\phi_1) e^{\mi n \Omega t}
\end{align}
\end{widetext}

As we discuss below, the TESS signal comprises of a set of peaks, with each term $H_{\a,\b}(t,\Omega)$ contributing a peak at a different time. The amplitudes and locations of these peaks contain information about the wakefield. A TESS signal can be calculated at several different spatial positions, giving spatial profiles of the wakefield's amplitude and frequency.

\subsection{Non-Gaussian Pulses}

So far we have made no assumptions about the probe and reference pulses, other than that they have well behaved Fourier transforms. In previous work \cite{Matlis2016} it was assumed that the incident probe and reference pulses are identical Gaussian pulses with no 3rd or higher order spectral phase: $\mathcal{E}_\mathrm{pr}(\omega)=\mathcal{E}_\mathrm{r}(\omega)=\mathcal{E}_0(\omega) \equiv A e^{-\frac{1}{2}(1+\mi \sigma) \left( \frac{\omega-\omega_0}{\delta \omega}\right)^2 }$. Here we remove the restriction on the spectral shape of the probe and reference pulses, whilst retaining the approximation that third and higher order phases can be neglected. Hence we may write:

\begin{align}
\mathcal{E}_\a (\omega) = \left| \mathcal{E}_\a (\omega) \right| \exp & \left \lbrace \mi \left[ \psi_\a^{(0)} + \psi_\a^{(1)} \left( \omega - \omega_0 \right) + \right. \right. \nonumber \\*
& ~~~ \left. \left. \frac{1}{2}\psi_\a^{(2)} \left( \omega - \omega_0 \right)^2 + \ldots \right ] \right \rbrace,
\end{align}
where the first order spectral phase $\psi_\a^{(1)}$ is the group delay, and the second order spectral phase $\psi_\a^{(2)}$ is the group delay dispersion (GDD). The group delay describes the arrival time of the central frequency $\omega_0$ before the delay $\Delta t$ has been introduced; for our incident probe and reference pulses $E_\mathrm{pr}(\zeta)$ and $E_\mathrm{r}(\zeta)$ we have defined $\psi_\mathrm{pr}^{(1)}=\psi_\mathrm{r}^{(1)}$ without loss of generality.

It can be shown (see Appendix~\ref{sec: TESS Peak}) that, if the difference in the 2nd order spectral phase of the probe and reference pulses is sufficiently small, $\psi_{\r}^{(2)} \approx \psi_{\p\r}^{(2)} \approx \psi^{(2)}$, the function $H_\mathrm{pr,r}(t,\Omega)$ peaks at $t_\Omega =  \psi^{(2)} \Omega + \psi_\r^{(1)} - \psi_{\p\r}^{(1)} = \psi^{(2)} \Omega$, at which time the peak amplitude is described by a cross-correlation:

\begin{align}
\left| H_\mathrm{pr,r}(t_\Omega,\Omega) \right| = \frac{1}{\sqrt{2\pi}} \int^{\infty}_{-\infty} \left| \mathcal{E}_\mathrm{pr}(\omega  - \Omega ) \right| \left| \mathcal{E}_\mathrm{r}(\omega) \right| \d \omega
\label{eqn: TESS Peak Amplitude}
\end{align}

From eqns~\eqref{eqn: TESS Signal} and~\eqref{eqn: TESS Peak Amplitude} we see that in general the TESS signal is composed of the DC peak at $t=0$, the two familiar delay sidebands at $t_0=\pm \Delta t$, and a series of equally spaced satellites either side of the delay sidebands, with the $k$-th order satellite of the $t_0=\Delta t$ peak located at $t_k = \Delta t + k \omega_{\p0} \psi^{(2)}$, see Fig.~\ref{fig: TESS Schematic}. This gives us a measurement of the wakefield frequency, using $\omega_\mathrm{p0} = (t_k-\Delta t)/ k \psi^{(2)}$, and hence an estimate of the electron density of the plasma, as $n_{e0}=\omega_\mathrm{p0} m_e \epsilon_0 /e^2$.

Additionally, the height of the satellite relative to that of the sideband is described by:

\begin{align}
r_k = \frac{ J_k(\phi_1 ) }{ J_0(\phi_1 ) } \mathcal{F}(k \omega_{\p0}), \label{Eqn: r for N=1}
\end{align}
where the spectral overlap factor $\mathcal{F}$ is given by:

\begin{align}	
	\mathcal{F}(\Omega) &\equiv  \frac{\int_{-\infty}^\infty \left|\mathcal{E}_\mathrm{pr}(\omega - \Omega) \right| \left| \mathcal{E}_\mathrm{r}(\omega) \right|\d \omega}{\int_{-\infty}^\infty \left|\mathcal{E}_\mathrm{pr}(\omega) \right| \left| \mathcal{E}_\mathrm{r}(\omega) \right|\d \omega} \nonumber \\
	&= \frac{\int_{-\infty}^\infty \sqrt{I_\mathrm{pr}(\omega - \Omega) I_\mathrm{r}(\omega)}\d \omega}{\int_{-\infty}^\infty \sqrt{I_\mathrm{pr}(\omega) I_\mathrm{r}(\omega)}\d \omega} , \label{Eqn: F_in_general}
\end{align}
where $I_{\p\r}$ and $I_\r$ are the spectral intensities of the \emph{input} probe and reference pulses. For probe and reference pulses of a bandwidth $\delta \omega$, we would expect this spectral overlap factor to become important when $\omega_\mathrm{p0} \gtrsim \delta \omega$.

For the case of identical Gaussian probe and reference pulses with $\sqrt{I(\omega)} = \left| \mathcal{E}_0(\omega) \right| = A e^{  - \frac{1}{2}\left( \frac{\omega-\omega_0}{\delta \omega} \right)^2 }$ we find:
\begin{align}	
	f(k\Omega) \equiv \mathcal{F}_\mathrm{Gauss}(k \Omega) = \exp\left[ - \left(\frac{1}{4}\frac{ k \Omega}{\delta \omega}\right)^2\right], \label{Eqn: F_for_Gaussian}
\end{align}
which agrees with the result found by Matlis et al\cite{Matlis2016}.

\begin{figure*}[h!]
\begin{center}
\includegraphics[scale=1.0]{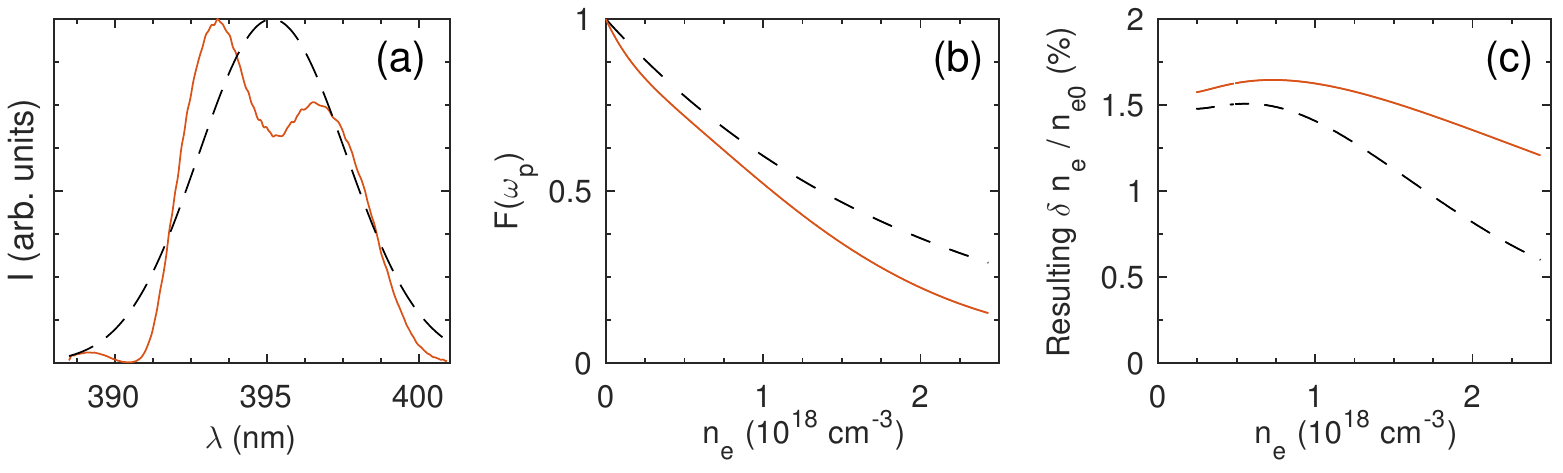}
\caption{a) The measured spectrum of the probe pulse in a recent experiment \cite{Cowley2017} (solid red line) and a Gaussian fit to it (dashed black line). b) Comparison of the spectral overlap factors $\mathcal{F}(\omega_{\p0})$ (solid, red) and $\mathcal{F}_\mathrm{Gauss}(\omega_{\p0})$ (dashed, black) evaluated at the first order TESS peak as a function of gas cell pressure, assuming that the hydrogen gas was fully ionized by the driving laser. c) Deduced wakefield amplitude as a function of cell pressure assuming spectral overlap factors $\mathcal{F}(\omega_{\p0})$ (solid, red) and $f(\omega_{\p0})$ (dashed, black), showing the mis-measurement of an example wakefield amplitude when assuming a Gaussian profile.}
\label{fig: Bandwidth Factor F}
\end{center}
\end{figure*}

Eqn~\eqref{Eqn: F_for_Gaussian} can lead to inaccurate determination of the wakefield amplitude for practical probe and reference pulses. For example, in recent experiments \cite{Cowley2017} we measured low amplitude wakefields $\delta n_e / n_{e0} \sim 1 \%$ using frequency doubled probe and reference pulses, the spectra of which were far from Gaussian, as shown in Fig.~\ref{fig: Bandwidth Factor F}a). The spectrum has a bandwidth of approximately $\delta \omega \approx 40$ rad ps$^{-1}$ and therefore a density of only {$n_e \approx 0.5 \times 10^{18}$ cm$^{-3}$} produces a plasma frequency $\omega_{\p0} \approx \delta \omega$. As a consequence, even at low electron densities the Gaussian approximation $\mathcal{F}_\mathrm{Gauss}(k\omega_{\p0})$ used by Matlis et al \cite{Matlis2016} diverges from the real overlap factor, making it unsuitable for accurately measuring wakefield amplitudes when the probe spectrum is non-Gaussian. For example, at a density of $n_e \approx 2.5 \times 10^{18}$ cm$^{-3}$, the Gaussian approximation reduced the calculated wakefield amplitude by a factor of approximately two.

Accurately measuring the wakefield frequency and amplitude over a range of experimental conditions can allow us to produce a resonance curve of wakefield amplitude with plasma frequency. In our recent experiments \cite{Cowley2017} this allowed us to infer information about the laser pulse which is driving the wakefield. For instance, for laser drivers with $a_0 \ll 1$ we expect the wakefield amplitude to increase linearly with the intensity of the drive pulse. Measuring violations of this trend would lead us to infer both an intense laser driver, with $a_0 \sim 1$, and a non-linear wakefield. We discuss estimating the wakefield amplitude under these conditions in section~\ref{sec: Non-linear Wakefields}.

\subsection{Non-Equal GDDs}

In general, the probe and reference pulses will have different  GDDs; this difference, which is usually small, arises from differences in the materials present in their optical paths. Whereas the previous work by Matlis et al \cite{Matlis2016} has ignored these effects, they may cause changes in the locations of the satellites, and hence on the deduced plasma frequency. Using the Fourier shift theorem it can be shown that $H_\mathrm{pr,r}(t, k \omega_{\p0}) = H_\mathrm{r,pr}^*(-t, - k \omega_{\p0}) e^{\mi k \omega_{\p0} t}$ and the satellite peak location must depend equally on the probe and reference pulses. We show in Appendix~\ref{sec: Non-Equal GDDS} that for the case of non-identical Gaussian pulses the satellite peak is located at $t = \Delta t + k \omega_{\p0} \psi^{(2)}_\mathrm{eff}$, where $\psi^{(2)}_\mathrm{eff}$ is the mean of the probe and reference GDDs, weighted by the square of their bandwidths:

\begin{align}
\psi^{(2)}_\mathrm{eff} &=  \frac{ \delta \omega_\mathrm{pr}^2 \psi^{(2)}_\mathrm{pr} +  \delta \omega_\mathrm{r}^2\psi^{(2)}_\mathrm{r} } {\delta \omega_\mathrm{pr}^2 + \delta \omega_\mathrm{r}^2} 
\end{align}

In the case where probe and reference pulses have the same bandwidth $\delta \omega_\mathrm{pr} \approx \delta \omega_\mathrm{r}$, this reduces to the arithmetic mean GDD $\psi^{(2)}_\mathrm{eff} = \frac{1}{2}\left[ \psi^{(2)}_\mathrm{pr} +  \psi^{(2)}_\mathrm{r} \right]$. In previous experimental work \cite{Cowley2017} we measured a difference in GDDs of around $1000$ fs$^2$ compared to $\psi_\mathrm{eff}^{(2)} \approx 20,000$ fs$^2$. In this case, using the probe GDD alone and assuming $\psi^{(2)}_\mathrm{eff} \approx \psi^{(2)}_\mathrm{pr}$ would result in an error in the calculated plasma frequency of around $2\%$.

\section{Non-linear Wakefields} \label{sec: Non-linear Wakefields}

\subsection{Quasi-Linear Wakefields}

\begin{figure*}[h!]
\begin{center}
\includegraphics{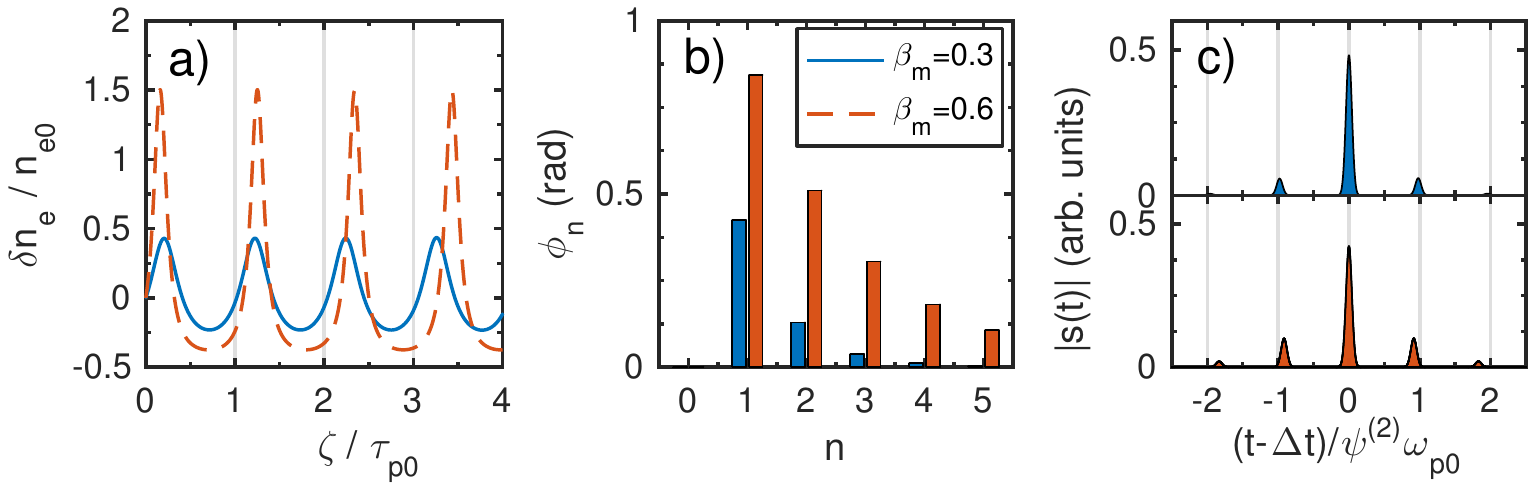}
\caption{a) Calculated temporal behaviour of the relative density of quasi-linear plasma waves for ${\beta_\mathrm{m}=0.3}$ (blue) and $\beta_\mathrm{m}=0.6$ (red). b) Harmonic amplitudes $\phi_n$ of the phase shift experienced by a 400 nm probe pulse co-propagating for 1 mm with the plasma waves shown in a). c) Calculated TESS signals for the plasma waves in (a) and the phase shifts shown in (b) at $\beta_\mathrm{m}=0.3$ (blue) and $\beta_\mathrm{m}=0.6$ (red).}
\label{fig: nonLinear TESS}
\end{center}
\end{figure*}

We now consider the extension of TESS to the characterization of non-linear plasma wakefields, where electrons in the plasma wave have relativistic velocities. Following Akhiezer and Noble \cite{Akhiezer1956, Noble1985}, in the quasi-linear regime the plasma wave amplitude can be described by the maximum electron velocity through $\beta_\mathrm{m} = v_\mathrm{e, max} / c$, where ${\delta n_e} / { n_{e0}} = {\beta} / {(1-\beta)}$ and $|\beta| \leq \beta_\mathrm{m}$. The plasma waves are linear in the regime $\beta_\mathrm{m} \ll 1$. Fig.~\ref{fig: nonLinear TESS}a) shows the calculated relative wave amplitude $\delta n_e / n_{e0}$ as a function of the co-moving coordinate $\zeta$ for two values of $\beta_\mathrm{m}$ at a plasma density of $n_{e0}=10^{18}$ cm$^{-3}$. It can be seen that as $\beta_\mathrm{m}$ increases, the wave becomes more sharply peaked and the period of the plasma wave is increased as $\tau_\p = \sqrt{\bar{\gamma}} \tau_{\p0}$, where $\bar{\gamma}$ is the normal relativistic factor averaged over one cycle.

As the wakefield remains both periodic, with a period $\tau_\p$, and continuous, the phase change acquired by the probe can be decomposed into a linear combination of harmonics: {$\phi_\mathrm{wake}(\zeta) \approx \sum_{n=1}^N \phi_n \sin(n \omega_\p \zeta + \theta_n)$}, where $\omega_\p = \omega_{\p0}/\sqrt{\bar{\gamma}}$ and $N$ can be arbitrarily large. Fig.~\ref{fig: nonLinear TESS}b) shows the amplitudes $\phi_n$ of the first five harmonics resulting from a 400 nm probe co-propagating over a distance of 1 mm with the plasma waves shown in Fig.~\ref{fig: nonLinear TESS}a). Whereas at $\beta_\m=0.3$ the phase change is dominated by the fundamental $n=1$, as $\beta_\m$ increases the wave becomes more non-linear and the higher harmonics at $n>1$ become relatively more significant. The Jacobi-Anger expansion can be applied to each of these $N$ harmonics as in the case of a linear wakefield, and so the electric field of the transmitted probe can be expanded as,

\begin{widetext}
\begin{align}
E_\mathrm{pr}' (\zeta) &= E_\mathrm{pr}(\zeta) \prod_{n=1}^N e^{\mi \phi_n \sin (n \omega_\p \zeta + \theta_n)} \nonumber \\
 &= E_\mathrm{pr}(\zeta) \prod_{n=1}^N \left[ \sum_{k_n=-\infty}^\infty J_{k_n}(\phi_n) ~ e^{\mi k_n \left( n \omega_\p \zeta + \theta_n \right)} \right] \nonumber \\
 &= E_\mathrm{pr}(\zeta) \sum^\infty_{k_1=-\infty} \ldots \sum^\infty_{k_N=-\infty} J_{k_1}(\phi_1) \ldots J_{k_N}(\phi_N) e^{\mi \left( \sum_{n=1}^N k_n n \right) \omega_\p \zeta } e^{ \mi \sum_{n=1}^N k_n \theta_n} \\
\mathcal{E}_\mathrm{pr}'(\omega) &= \sum^\infty_{k_1=-\infty} \ldots \sum^\infty_{k_N=-\infty} J_{k_1}(\phi_1) \ldots J_{k_N}(\phi_N) e^{ \mi \sum_{n=1}^N k_n \theta_n } \mathcal{E}_\mathrm{pr} \left( \omega -  \sum_{n=1}^N k_n n~ \omega_\p \right).
\end{align}
\end{widetext}

The transmitted probe spectrum therefore contains contributions from the $N$ (in principle, infinite) harmonics, each of which in turn produces an infinite set of carrier waves, labelled by the integers $k_n$, which can be both positive or negative. By truncating the phase expansion at finite $N$, each term is uniquely labelled with a set of $N$ different indices, from $k_1$ to $k_N$. However, as $|k_n|$ increases, the amplitude of the carrier wave decreases since $|J_{k_n}(\phi_n)| \rightarrow 0$ as $|k_n| \rightarrow \infty$. For any finite $\phi_n$ the carrier wave expansion can therefore be terminated at a finite value of $|k_n| \leq K$. The larger the phase $\phi_n$, the larger $K$ must be for accurate reconstruction of the wakefield. Every term in this expansion can therefore be described with a point on a discrete $N$-dimensional grid with $2K+1$ points on each side, although both $K$ and $N$ can be arbitrarily large; the point is described by $\mathbf{k} = (k_1,k_2,k_3,k_4,\ldots, k_N)$. There are $(2K+1)^N$ such points in the grid. The reference pulse is the same as before, and the interference between the two gives rise to a TESS signal. We shall consider only the term corresponding to the peaks near $t=\Delta t$, which arises from the interference term $\mathcal{E}_\mathrm{pr}'(\omega) \mathcal{E}_\mathrm{r}'^*(\omega)$, i.e.,

\begin{widetext}
\begin{align}
s(t) &\approx \ldots +  \sum_{k_1=-K}^K \ldots \sum_{k_N=-K}^K J _{k_1}(\phi_1) \ldots J_{k_N}(\phi_N) e^{\mi \sum_{n=1}^N k_n \theta_n} H_\mathrm{pr,r} \left( t-\Delta t,\sum_{n=1}^N k_n n \omega_\p \right) + \ldots \nonumber \\
&= \ldots +  \sum_{\kappa=-K}^{K} Z_\kappa \left( \lbrace \phi_n \rbrace, \lbrace \theta_n \rbrace \right) H_\mathrm{pr,r} \left( t-\Delta t,\kappa\omega_\p \right) + \ldots \label{Eqn: non-linear s(t)} \\
\textrm {where}  ~~~ Z_\kappa \left( \lbrace \phi_n \rbrace,\lbrace \theta_n \rbrace \right) &= \sum_{ \mathbf{k} \in S_\kappa} J_{k_1}(\phi_1) \ldots J_{k_N}(\phi_N) e^{ \mi \sum_{n=1}^N k_n \theta_n } \label{Eqn: non-linear Z} \\
\textrm{for a subset} ~~~ S_\kappa &= \left \lbrace \left ( k_1, \ldots , k_N \right) : \sum_{n=1}^N n k_n = \kappa \right \rbrace
\end{align}
\end{widetext}

In eqn~\eqref{Eqn: non-linear s(t)} we have divided the set of points $\mathbf{k}$ in the grid into subsets, each labelled with the new integer $\kappa$, which, as seen in eqn~\eqref{Eqn: non-linear s(t)}, is the effective harmonic order. Each of these subsets $S_\kappa$ comprises those points from the original $N$ dimensional grid which also lie upon a particular $N-1$ dimensional plane, which is perpendicular to the vector $\mathbf{v}=(1,2,3,\ldots,N)$. The new label $\kappa$ is then the plane number, $\kappa = \mathbf{v}\cdot \mathbf{k}$, which can be both positive or negative. For instance, the $\kappa = 2$ subset contains points $\mathbf{k} = (2,0,0,0,\ldots)$, $(0,1,0,0,\ldots)$, $(-1,0,1,0,\ldots)$ and $(2,-2,0,1,\ldots)$, amongst many others. To completely span the original grid we must consider all planes up to $|\kappa| \leq \kappa_\mathrm{max} = K + 2K + \ldots + NK = KN(N+1)/2$.

The TESS signal $s(t)$ now looks very similar to that for a linear wakefield, with the substitution of a new complex number $Z_\kappa$, which includes contributions from all of the wakefield's frequency components, instead of the Bessel function $J_k$, which only accounts for the fundamental frequency. The crucial part of this signal again consists of a sideband peak at $t_0 = \Delta t$ and a series of satellite peaks spaced around it at times $t_\kappa = \Delta t + \kappa \psi^{(2)} \omega_\p$. For a quasi-linear wakefield, however, each peak will contain contributions from many different frequency components; the $\kappa$-th order peak will have contributions from all terms in $S_\kappa$, including all harmonics such that $\sum_{n=1}^N n k_n = \kappa$. 

Fig.~\ref{fig: nonLinear TESS}c) shows the TESS satellites resulting from the phase shifts shown in Fig.~\ref{fig: nonLinear TESS}b), for quasi-linear plasma waves with $\beta_\m=0.3$ and $\beta_\m=0.6$. While the relative heights of the satellite peaks are slightly greater for $\beta_\m=0.6$, the form of the signal is largely unchanged from that of a linear wakefield. The satellites are located slightly closer to the origin because the plasma frequency is reduced as $ \omega_\p = \omega_{\p0} / \sqrt{\bar{\gamma}}$. The satellite peak heights, relative to the sideband, are now:

\begin{align}
r_\kappa = \left | \frac{Z_\kappa \left( \lbrace \phi_n \rbrace, \lbrace \theta_n \rbrace \right)}{Z_0 \left( \lbrace \phi_n \rbrace, \lbrace \theta_n \rbrace \right)} \right | \mathcal{F}\left( \kappa \omega_\p \right) . \label{Eqn: non-linear ratios}
\end{align}

In general, the wakefield is difficult to recover from measured $r_\kappa$, as every combination of $\left( k_1, k_2, \ldots , k_N \right) \in S_\kappa$ must be accounted for. This set, the points of an $N$ dimensional grid of side length $2K + 1$ which also lie on the $(N-1)$ dimensional plane $\kappa = \mathbf{v}\cdot \mathbf{k}$, contains on the order of $|S_\kappa|\sim (2K + 1)^{(N-1)}$ members and the contribution of each must be calculated. For example, limiting the set to $N=5, K=5$ gives $|S_2|=2583$ solutions to $\kappa=2$.

Once the sets $S_0$ and $S_\kappa$ have been found for each of $N$ satellite peaks, obtaining the wakefield profile from the measured peak heights would involve measuring the amplitude and phase of $N$ satellite peaks, yielding $2N$ non-linear simultaneous equations, and solving for the $2N$ unknowns $\lbrace \phi_n \rbrace$ and $\lbrace \theta_n \rbrace$. In practice, however, it is difficult to measure more than a few satellite peaks since the overlap factor $\mathcal{F}(\kappa \omega_\p)$ becomes small as $\kappa$ increases. Given a set of experimental parameters and a particular model of the wake, such as Akhiezer and Noble quasi-linear plasma waves, it is nonetheless possible to tackle this problem numerically by calculating the values $\lbrace \phi_n \rbrace$ and $\lbrace \theta_n \rbrace$ for a range of wakefield amplitudes. Instead of $2N$ unknowns the wakefield is described by a single amplitude $\beta_\mathrm{m}$, and using $\lbrace \phi_n \rbrace$ and $\lbrace \theta_n \rbrace$, a lookup table of ratios $|Z_\kappa / Z_0|$ against wakefield amplitudes $\beta_\mathrm{m}$ can be constructed. Comparing measured ratios to these values then yields the wakefield amplitudes. In the following section we demonstrate this approach for simulated data.

\begin{figure*}[h!]
\begin{center}
\includegraphics{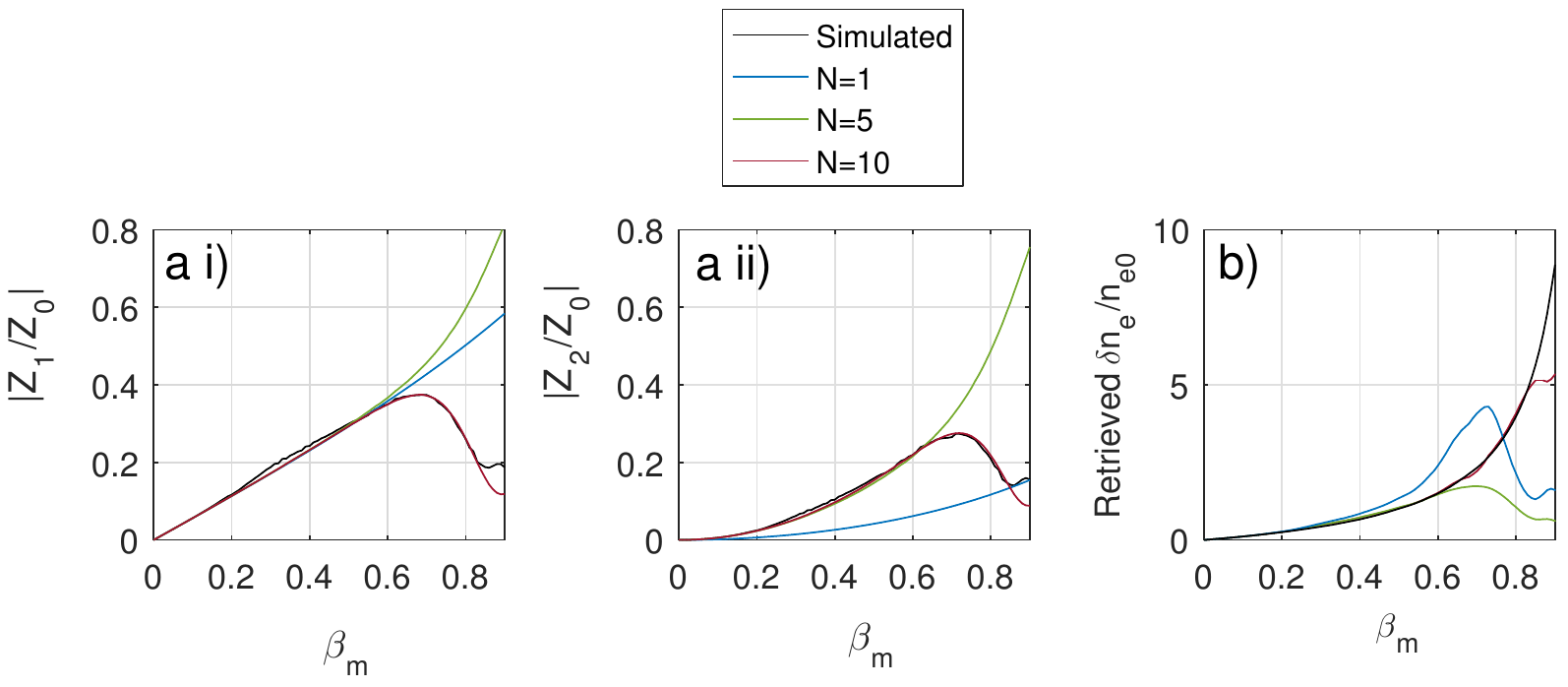}
\caption{a) The calculated ratios $|Z_\kappa/Z_0|$ against plasma wave amplitude $\beta_m$ for i) the $\kappa=1$ and ii) $\kappa=2$ satellite peaks. The ratios are approximated using $N=1$ (blue), $N=5$ (green) and $N=10$ (red) and are compared to simulations of the ideal ratios (black). b) The wakefield amplitude retrieved using the peak ratios $|Z_\kappa/Z_0|$ against the simulated plasma wave amplitude.}
\label{fig: nonLinear TESS ratios}
\end{center}
\end{figure*}

\subsubsection{Simulated TESS Analysis}

In order to demonstrate this procedure, the TESS signals, similar to those shown in Fig.~\ref{fig: nonLinear TESS}c), were calculated for simulated plasma waves with maximum electron velocity $\beta_\mathrm{m}$ of up to $0.9$, using the same probe frequency, plasma density and interaction length as above. From these TESS signals the peak ratios $r_1$ and $r_2$ were extracted and divided by the overlap factors to calculate $|Z_1/Z_0|$ and $|Z_2 / Z_0|$. These simulated ratios were compared to expectations from Eqn.~\eqref{Eqn: non-linear Z}, where $Z_0$, $Z_1$ and $Z_2$ were calculated for cold 1D plasma waves as above. In all cases we limited the Bessel expansion to third order, $|k_n|\leq 3$ as $J_4(x)/J_3(x) < 0.13$ for $x<1$. We considered three possible analyses, with truncation of the harmonic expansion at: N = 1, corresponding to assumption of a linear wakefield; N = 5, for which there are around $|S_\kappa| ~ 400$ combinations; and N = 10, for which $|S_\kappa| ~ 3$ million. The results, shown in Fig.~\ref{fig: nonLinear TESS ratios}a), demonstrate that as wakefield amplitude increases more harmonics must be accounted for in Eqn.~\eqref{Eqn: non-linear Z} in order to correctly calculate the ratios $|Z_1/Z_0|$ and $|Z_2 / Z_0|$. Whereas the linear approximation $N=1$ fails for $\beta_\mathrm{m} > 0.1$, using $N=10$ the simulations match the calculations very closely up to a wakefield amplitude of around $\beta_\mathrm{m} \approx 0.8$.

Next, these ratios were used to calculate lookup tables of $|Z_1/Z_0|$ and $|Z_2 / Z_0|$ over different wakefield amplitudes; each value of $N$ gave a different lookup table. The wakefield amplitude of the simulated TESS spectra was then estimated by using these lookup tables and the known interaction length and wakefield frequency. For each simulated TESS spectra we chose the value of $\beta_\mathrm{m}$ which minimised the distance (or 2-norm) between the measured ratios and the ratios on the lookup table. For $N=1$ we ignored $Z_2$ and followed the same procedure as for linear plasma waves. As shown in Fig.~\ref{fig: nonLinear TESS ratios}b), the linear assumption used in section~\ref{sec: General Probe Pulses} works surprisingly well for non-linear plasma waves, and the accuracy can be increased by further expanding $N$. The linear TESS procedure, $N=1$, begins to diverge from the true value at wakefield amplitudes of $\beta_\mathrm{m} \approx 0.2$, or $\delta n_e / n_{e0} \approx 25\%$, and subsequently overestimates the wakefield amplitude by around $10 \%$ of the true value. Using the $N=5$ expansion, however, accurately measures wakefield amplitudes up to $\beta_\mathrm{m} \approx 0.6$, and the expansion to $N=10$ only fails at $\beta_\mathrm{m} > 0.8$.

There is therefore a trade off between accuracy and computation time. We have demonstrated that an expansion to the 10th harmonic of the plasma frequency allows us to accurately calculate the amplitude of a quasi-linear wakefield up to $\beta_\mathrm{m} \approx 0.8$. However, it requires calculating the sum of around 3 million components for each satellite peak at each wakefield amplitude, which takes several minutes on a desktop computer. When truncating the expansion to $N=5$ harmonics, however, the process requires only around 400 components and takes less than 0.1 seconds. These considerations are important as each lookup table is only valid for a certain experimental set up, with a given plasma density, probe frequency and interaction length. Once the lookup table is calculated, however, retrieval of the wakefield amplitude and frequency is extremely fast. 

The analysis can be simplified in two cases. If the wakefield is not too non-linear, $\beta_\mathrm{m} \ll 1$, the original frequency component at $\omega_\p$ dominates, such that ${\forall n>1 (\phi_n \ll \phi_1)}$. In this case, it is possible to truncate the expansion at $N=1$ and use the same procedure as for linear wakefields, with $r_1 \approx \left[{J_1 (\phi_1)} / {J_0 (\phi_1)}\right] \mathcal{F}(\omega_\p) $. This sacrifices only a small amount of accuracy, on the few percent level. On the other hand, if the phase change is sufficiently small, $\phi(\zeta) \ll 1$, such as at low densities or short interaction lengths, it is feasible to limit the order of the Bessel peaks $|k_n| \leq 1$, as $J_2(x)/J_1(x)<0.13$ for $x<0.5$. If we further allow only one $k_n$ to be non-zero, the ratio of heights of the $\kappa$-th satellite to the sideband is much easier to calculate:
\begin{align}
r_\kappa = \frac{ J_1(\phi_\kappa) }{ J_0(\phi_\kappa) } \mathcal{F}(\kappa\omega_\p)
\end{align}

This situation can always be achieved for a given experiment by reducing the interaction length until ${\phi(\zeta) \ll 1}$. As with conventional interferometry, phase changes which are too large make reconstruction difficult, but small phase changes are difficult to measure.

\subsection{General Wakefields}

In general, wakefields need not be periodic, and hence contain many frequencies which aren't multiples of the plasma frequency. If we approximate this with a finite set of $N$ frequencies, which are not uniformly spaced, the phase change due to the wakefield can be written $\phi_\mathrm{wake}(\zeta) = \sum_{n=1}^N \phi_n \sin(\omega_n \zeta + \theta_n)$. By comparison with the results for a quasi-linear wakefield with $n \omega_\p\rightarrow \omega_n$ we can find the resulting TESS signal:

\begin{align}
s(t) = \ldots + & \sum_{k_1=-\infty}^\infty \ldots \sum_{k_N=-\infty}^\infty J_{k_1}(\phi_1) \ldots J_{k_N}(\phi_N) \nonumber \\*
& \cdot  e^{\mi \sum_{n=1}^N k_n \theta_n} H_\mathrm{pr,r} \left( t-\Delta t,\sum_{n=1}^N k_n \omega_n \right) + \ldots
\end{align}

While this looks very similar to the result for quasi-linear wakefields, the peaks for general wakefields will lie either side of the original TESS satellites at locations $t = \Delta t + \psi^{(2)} \left( \sum_{n=1}^N k_n \omega_n \right)$. These positions are not equally spaced and therefore each of the TESS satellites will be split into new peaks. However, as one of the sinusoidal components is at the plasma frequency, $\omega_1=\omega_\p$, there will still be peaks at $t_k = \Delta t + k \psi^{(2)} \omega_\p$, from the cases $\lbrace k_n \rbrace = \lbrace k,0,0,\ldots \rbrace$

\begin{figure*}[h!]
\begin{center}
\includegraphics{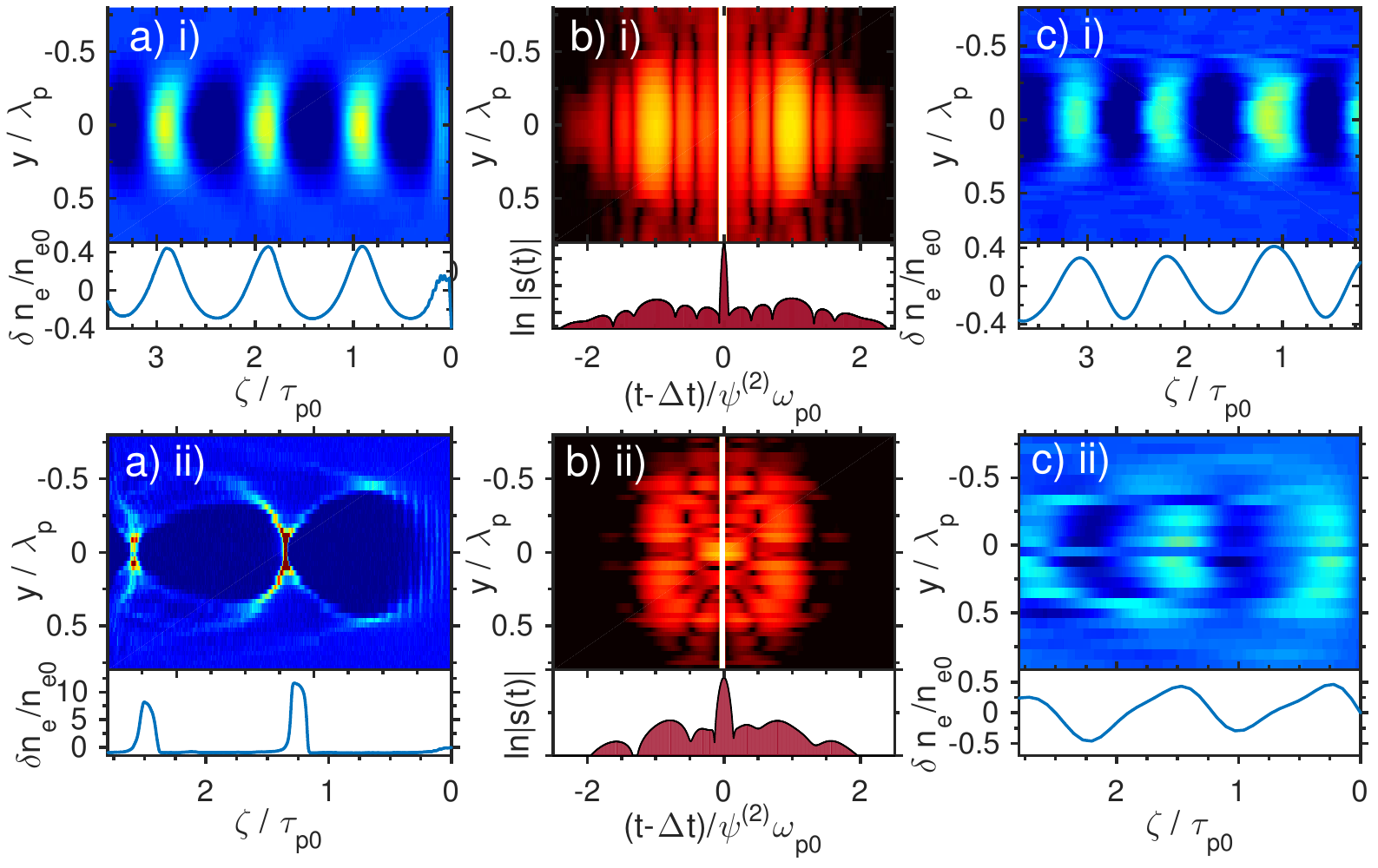}
\caption{a) Simulated electron density maps of a wakefield in i) a weakly non-linear regime and ii) the strongly non-linear bubble regime. b) Maps of the calculated TESS signal resulting from these density profiles, plotted in both space and in time against the expected peak positions. The magnitude of the TESS signal is plotted on a logarithmic scale. c) A reconstruction of the electron density map of the wakefield using the TESS analysis, using $N=4$ peaks from the TESS signal. In all plots the profile on-axis is shown below.}
\label{fig: Bubble Regime TESS}
\end{center}
\end{figure*}

This is demonstrated for example simulated wakefields, in i) a weakly non-linear regime and ii) the bubble regime, shown in Fig.~\ref{fig: Bubble Regime TESS}. In the bubble regime the ponderomotive force from an intense laser pulse is sufficient to completely evacuate a region of electrons, leaving only ions within a bubble with a diameter equal to the plasma period. The density maps were simulated with the EPOCH Particle-In-Cell code \cite{Arber2015}, with a bi-Gaussian drive laser with a peak beam intensity of i) $2 \times 10^{18}$~Wcm$^{-2}$ and ii) $4 \times 10^{19}$~Wcm$^{-2}$, passing through a high density plasma at $n_e = 10^{19}$~cm$^{-3}$. The pulse was matched to the plasma wavelength and period with a $w_0 = 3$ $\mu$m spot and a duration of $t_\mathrm{FWHM}=13$ fs. The resulting density profiles are neither sinusoidal nor periodic and hence will contain many frequency components. The resulting TESS signals in both space and time were simulated for identical and Gaussian probe and reference pulses with $400$ nm wavelength and $10$ fs bandwidth limited duration, each stretched to $1$ ps FWHM duration.

Fig.~\ref{fig: Bubble Regime TESS}a) shows the density maps in the two cases: in the weakly non-linear case several plasma periods are captured by the simulation, whereas in the bubble regime only two bubbles are shown, separated by $\tau_\p$. Because the wakefield is strongly non-linear the electrons are relativistic and the plasma period is longer than expected for the density, $\tau_\p = \sqrt{\bar{\gamma}} \tau_{\p0}$, giving $\omega_\p=\omega_{\p0}/\sqrt{\bar{\gamma}}$. Fig.~\ref{fig: Bubble Regime TESS}b) shows the TESS signal for each spatial position of the simulation, with many peaks present at different spatial positions. The 1st and 2nd order satellites can be seen at locations $t_1 \approx \Delta t + 0.8\, \omega_{\p0} \psi^{(2)}$ and $t_2 \approx \Delta t + 1.6\, \omega_{\p0}  \psi^{(2)}$, which implies that $\bar{\gamma} \approx 1.6$. From this it is possible to estimate the wakefield amplitude as $\bar{\beta}\approx 0.8$, or $\delta n_e / n_{e0} \approx 4$, which is a substantial underestimate.

Recovering the wakefield amplitude more accurately is difficult as in general many frequencies are required to completely reconstruct the wakefield; the amplitude of the fundamental at a frequency $\omega_\p$ is much smaller than the total height of the density fluctuation. Unlike periodic quasi-linear wakefields, there is no way to know from the satellite peak heights alone whether the amplitude is reduced because of the shape of the wakefield or because there are only a few plasma periods present. With only two plasma periods present in the strongly non-linear regime, applying the linear analysis of section~\ref{sec: General Probe Pulses} to the TESS signal in Fig.~\ref{fig: Bubble Regime TESS}b) measures the density fluctuation as $\delta n_e / n_{e0} \approx 0.9$, which is a factor of 10 lower than the true value. In the weakly non-linear regime, however, the wake amplitude is estimated as $\delta n_e / n_{e0} \approx 0.4$, close to the true value. In this situation the first TESS peak is sufficient for a first order approximation.

Fig.~\ref{fig: Bubble Regime TESS}c) shows attempted reconstructions by instead measuring the location, amplitude and phase of all peaks up to $t=\Delta t + 3 \psi^{(2)} \omega_{\p0}$ at each spatial position of the TESS spectrum. The frequency of each component was deduced as $\omega_n = (t_n - \Delta t) / \psi^{(2)}$, where $t_n$ was the location of the $n$-th peak; the amplitude was deduced from $r_n = \left[ J_1(\phi_n)/J_0(\phi_n) \right] \mathcal{F}(\omega_n)$, where $r_n = |s(t_n)/s(\Delta t)|$ is the measured ratio of peak heights; and the phase was deduced from $\theta_n \approx \arg \left[ s(t_n)/s(\Delta t) \right]$. This allowed us to reconstruct the phase change due to the wakefield as $\phi_\mathrm{wake} \approx \sum_{n=1}^N \phi_n \sin ( \omega_n \zeta + \theta_n )$, for each spatial position, and hence to reconstruct the density map of the wakefield.

While some of the general features of the wakefield are reproduced, only a limited number of peaks ($N=4$) were captured in the TESS signal and so this reconstruction inevitably fails to reproduce the true profile present in the simulation, particularly small spatial features. Whereas for the weakly non-linear wakefield it is still possible to reconstruct the density with reasonable accuracy, for the simulation in the bubble regime the reconstruction fails and the calculated wakefield amplitude is more than an order of magnitude too small. The accuracy is limited by the number of peaks that can be captured in the TESS signal, which in turn is effectively limited by the bandwidth of the probe pulse and the overlap factor $\mathcal{F}(\omega_n)$. Sharp and non-periodic features with durations less than the bandwidth limited duration of the probe pulse, for instance features such as an ionization front, are poorly reproduced by TESS, and are a source of noise when trying to extract the wakefield amplitude. Capturing shorter duration features requires using a shorter duration probe pulse with a broader bandwidth.

When attempting to reconstruct the density profile of a non-linear wakefield with a continuous range of frequency components, FDH is therefore likely to be more effective than TESS. We can see this by considering another approach to this problem, where we can approximate the phase with $N$ uniformly spaced frequencies separated by an arbitrarily small frequency, $\delta_\omega$. The phase can then be written as $\phi_\mathrm{wake}(\zeta) = \sum_{n=1}^N \phi_n \sin(n \delta_\omega \zeta + \theta_n)$, where many of the amplitudes $\phi_n$ are small or zero. This is a discrete Fourier transform (DFT), and approximates the true phase profile increasingly well at larger $N$ and smaller $\delta_\omega$. All of the information about the wakefield is contained in the region of the TESS signal surrounding the sideband at $t=\Delta t$, which can be written as:

\begin{align}
s(t) &= \ldots +  \sum_{\kappa=-\infty}^{\infty} Z_\kappa \left( \lbrace \phi_n \rbrace, \lbrace \theta_n \rbrace \right) H_\mathrm{pr,r} \left( t-\Delta t,\kappa \delta_\omega \right) + \ldots,
\end{align}
where as before $\kappa = \sum_{n=1}^N k_n n$ and $Z$ is defined in eqn~\eqref{Eqn: non-linear Z}. The potential TESS peaks are then separated by $\psi^{(2)} \delta_\omega$, although only some of these peaks will be non-zero. Fourier transforming this region of the TESS signal to the spectral domain yields:

\begin{align}
\mathcal{S}(\omega) &= \ldots +  \sum_{\kappa=-\infty}^{\infty} Z_\kappa \left( \lbrace \phi_n \rbrace, \lbrace \theta_n \rbrace \right) \nonumber \\*
& ~~~~~~~ \cdot \mathcal{E}_\mathrm{pr} \left( \omega - \kappa \delta_\omega \right) \mathcal{E}^*_\mathrm{r}\left( \omega \right) e^{- \i \omega \Delta t} + \ldots \\
&= \ldots + \mathcal{E}'_\mathrm{pr}(\omega) \mathcal{E}'^*_\mathrm{r}(\omega) + \ldots \nonumber
\end{align}

The reference pulse in the spectral domain is unchanged, but the presence of the wakefield has modulated the electric field of the probe, creating many copies of the probe pulse, spectrally shifted by multiples of $\delta_\omega$ and with amplitudes and phases described by $Z_\kappa$. The frequency interval $\delta_\omega$ can therefore be considered as the resolution of the DFT in the spectral domain. However, as the DFT becomes continuous with $\delta_\omega \rightarrow 0$, these copies, and the resulting satellite peaks in the TESS signal, are no longer distinct. In the continuum limit we cannot measure relative peak heights and the TESS procedure will inevitable fail. Instead we have isolated the spectral component $\mathcal{E}'_\mathrm{pr}(\omega) \mathcal{E}'^*_\mathrm{r}(\omega)$ and must continue the FDH procedure to reconstruct the phase due to the wakefield.

\subsubsection{Longitudinal Variation}

We have just discussed a situation where the wakefield is not periodic in the co-moving frame. The same phase change can arise, however, if the background plasma density varies longitudinally and the plasma frequency changes along the path of the probe pulse. By the nature of the experimental set up both FDH and TESS average the density profile along this path, with each point $\zeta$ in the co-moving frame corresponding to a line of points in space described by $z=v_g(\tau - \zeta)$. Whereas an FDH reconstruction risks obscuring the signal from one region by overlaying it with the signal from another region, TESS separates the density profile into its different frequency components. If the probe pulse encounters distinct regions of varying electron density along its path, such as in two stage injection-acceleration set-ups described in refs.~\cite{Malka2016,Golovin2015,Swanson2017}, the spectral interferogram will again contain components from several frequencies described by $\omega_n = \omega_\p(z_n)$.

As described above, this will have the effect of creating new peaks in the TESS spectrum, where each corresponds to the plasma frequency at a particular region along the path of the probe pulse. However, if we are aware of the longitudinal variation of the plasma density and have measured the length of each region, we can again look only at uniformly spaced peaks from the sets $\lbrace k_1, 0 , 0 \ldots \rbrace$, $\lbrace 0, k_2, 0 , 0 \ldots \rbrace$, $\lbrace 0, 0, k_3, 0 \ldots \rbrace$ etc. These relate to distinct regions along the path of the probe pulse, each with a distinct plasma density. So long as these do not overlap, if the density ramps between different regions are sufficiently short, it is possible to reconstruct the wakefield frequency and amplitude within each region as before, using:

\begin{align}
t_{x,k_x} &= \Delta t + k_x \psi^{(2)} \omega_{\p,x}, \\
r_{x,k_x} &= \frac{ J_{k_x}(\phi_x ) }{ J_0(\phi_x ) } \mathcal{F}(k_x \omega_{\p0}),  \\
\textrm{and   } \phi_x &= \frac{\omega_{\p,x}^2 L_x}{2 \omega_0 c} \frac{\delta n_e}{n_{e,x}},
\end{align}
where $L_x$ is the length of a region with plasma density $n_{e,x}$ and plasma frequency $\omega_{\p x}$. This causes a phase change of amplitude $\phi_x$ and peaks in the TESS spectrum at locations $t_{x,k_x}$.

\section{Conclusions}

We have extended the TESS analysis technique to probe and reference pulses of arbitrary temporal and spectral profile. This allows more accurate measurement of the frequency and amplitude of the wakefield in real situations by using the measured spectra of the probe and reference pulses instead of a Gaussian approximation. In turn this allows us to calculate the electron density of the plasma and to infer information about the laser pulse driving the wakefield.

In calculating wakefield amplitudes using TESS, the generalized spectral overlap factor given in eqn~\eqref{Eqn: F_in_general} can be calculated straightforwardly from the measured spectra of the probe and reference pulses. Using recent experimental results, we showed that the assumption of Gaussian probe and reference pulse spectra can lead to errors in the deduced amplitude of the plasma wakefield by a factor of around two. We have also demonstrated that when the probe and reference have different GDDs the TESS peak separation is described by the effective GDD, which is the mean of the probe and reference GDDs weighted by the square of their bandwidths.

We have also explored the applicability of TESS to measurements of non-linear relativistic plasma waves by decomposing the wakefield into harmonics of the plasma frequency. Simulations showed that the extension of TESS to quasi-linear plasma waves allowed accurate reconstruction of cold plasma waves with electron velocities as high as $\beta_\mathrm{m} \approx 0.8$. For high wakefield amplitudes finding the peak height ratios involves solving a Diophantine equation and summing over millions of contributions, but at lower amplitudes or smaller phase shifts the wakefield amplitude can be extracted rapidly. TESS therefore retains its advantages over FDH for wakefields in the quasi-linear regime. For general non-linear wakefields, however, it was only possible to measure the wakefield frequency and not the amplitude, and in this regime an FDH phase reconstruction is required. On the other hand, for wakefields in plasmas with distinct regions of different density TESS has the capability to extract the wakefield amplitude in each region separately, but only if the length of these regions are known.

The extension of TESS to quasi-linear plasma waves is particularly relevant for measuring strong wakefields generated at low plasma densities. As laser wakefield experiments attempt to increase the electron energy gain through increasing the interaction length, TESS provides a means of rapidly diagnosing problems with the wakefield on-shot, without requiring electron injection. We have previously demonstrated \cite{Cowley2017} that TESS can work effectively at densities of $n_e \sim 10^{18}$ cm$^{-3}$ and below, accurately measuring wakefields with a relative amplitude as small as $1\%$. By demonstrating that TESS can also be effective for plasma waves with density fluctuations on the scale of $\delta n_e \sim n_{e0}$, it is possible to envisage applying it to accurately measure large amplitude wakefields at electron densities below $n_e \sim 10^{17}$ cm$^{-3}$.

\begin{acknowledgments}
This work was supported by the UK Science and Technology  Facilities  Council  (STFC  UK)  [Grants No. ST/J002011/1, No. ST/P002048/1, and No. ST/
M50371X/1];  the  Helmholtz  Association  of  German Research centers [Grant No. VH-VI-503]; and Air Force Office of Scientific Research, Air Force Material Command, USAF [Grant No. FA8655-13-1-2141].
\end{acknowledgments}

%

\newpage

\onecolumngrid
\appendix

\section{Detailed TESS Calculations} \label{sec: TESS Peak}

We consider the probe and reference pulses to be non-identical and to be of arbitrary spectral profile $\mathcal{E}_\mathrm{x}(\omega ) = \left| \mathcal{E}_\mathrm{x}(\omega ) \right| \exp \left( \mi \psi_\mathrm{x} \right)$. In evaluating eqn~\ref{eqn: H in general} we will need to calculate the difference in spectral phase:

\begin{align}
	\Delta \psi_\mathrm{a,b} (\omega,\Omega) &\equiv \psi_\mathrm{a} (\omega - \Omega) - \psi_\mathrm{b} (\omega) \nonumber \\
	&= \left[ \psi^{(0)}_\mathrm{a} + \psi^{(1)}_\mathrm{a} (\omega - \omega_0 - \Omega) + \frac{1}{2} \psi^{(2)}_\mathrm{a} (\omega - \omega_0 -\Omega) ^2 + \ldots \right] \nonumber \\
	&~~~~ - \left[ \psi^{(0)}_\mathrm{b} +  \psi^{(1)}_\mathrm{b} (\omega - \omega_0) + \frac{1}{2} \psi^{(2)}_\mathrm{a}  (\omega - \omega_0 ) ^2 + \ldots \right] \nonumber \\
	&= \frac{1}{2} \left( \psi^{(2)}_\mathrm{a} - \psi^{(2)}_\mathrm{b} \right) (\omega - \omega_0 ) ^2 +  \left( \psi^{(1)}_\mathrm{a} - \psi^{(1)}_\mathrm{b}  - \psi^{(2)}_\mathrm{a} \Omega \right)  (\omega - \omega_0 ) \nonumber \\
	& ~~~~ + \left( \psi^{(0)}_\mathrm{a} - \psi^{(0)}_\mathrm{b} - \psi^{(1)}_\mathrm{a} \Omega + \frac{1}{2} \psi^{(2)}_\mathrm{a} \Omega ^2 \right) + \ldots \nonumber \\
	&= A (\omega - \omega_0 ) ^2 + B(\Omega)  (\omega - \omega_0 ) + C(\Omega) + \ldots \\
	\mathrm{for}~~~~ A &= \frac{1}{2} \left( \psi^{(2)}_\mathrm{a} - \psi^{(2)}_\mathrm{b} \right) \label{eqn: Difference in spectral phase} \\
	B &=\psi^{(1)}_\mathrm{a} - \psi^{(1)}_\mathrm{b}  - \psi^{(2)}_\mathrm{a} \Omega \\
	C &= \psi^{(0)}_\mathrm{a} - \psi^{(0)}_\mathrm{b} - \psi^{(1)}_\mathrm{a} \Omega + \frac{1}{2} \psi^{(2)}_\mathrm{a} \Omega ^2
\end{align}

\subsection{Case $\psi^{(2)}_\mathrm{a} = \psi^{(2)}_\mathrm{b}$}

When the two pulses have equal GDD the coefficient $A=0$ and hence:

\begin{align}
H_\mathrm{a,b}(t,\Omega) &\equiv \frac{1}{\sqrt{2\pi}}\int^{\infty}_{-\infty} \mathcal{E}_\mathrm{a}(\omega - \Omega) \mathcal{E}_\mathrm{b}^*(\omega )  \exp \left( \mi \omega t \right) \d \omega \nonumber \\
&= \frac{1}{\sqrt{2\pi}}\int^{\infty}_{-\infty} \left| \mathcal{E}_\mathrm{a}(\omega - \Omega) \right| \left| \mathcal{E}_\mathrm{b}(\omega ) \right| \exp \left[ \mi B(\Omega) (\omega - \omega_0)  + \mi C(\Omega) +  \mi \omega t \right] \d \omega \nonumber \\
&= \frac{1}{\sqrt{2\pi}} e^{\mi C(\Omega)} e^{- \mi \omega_0 B(\Omega)} \int^{\infty}_{-\infty} \left| \mathcal{E}_\mathrm{a}(\omega - \Omega) \right| \left| \mathcal{E}_\mathrm{b}(\omega ) \right| \exp \left\lbrace \mi \omega \left[ B(\Omega) + t \right]  \right\rbrace \d \omega \label{Eqn: H when A=0}
\end{align}

Since $ \left| \mathcal{E}_\mathrm{a}(\omega - \Omega) \right| \left| \mathcal{E}_\mathrm{b}(\omega ) \right|$ is real, this integral is maximised when the phase $psi=\omega \left[ B(\Omega) + t \right] $ is stationary with respect to $\omega$ at $t_\Omega = - B(\Omega) = \psi^{(2)}_\mathrm{a} \Omega + \psi^{(1)}_\mathrm{b} - \psi^{(1)}_\mathrm{a}$. At this peak, the integral has an amplitude:

\begin{align}
\left| H_\mathrm{a,b}(t_\Omega,\Omega) \right| = \frac{1}{\sqrt{2\pi}} \int^{\infty}_{-\infty} \left| \mathcal{E}_\mathrm{a}(\omega - \Omega) \right| \left| \mathcal{E}_\mathrm{b}(\omega ) \right| \d \omega \label{Eqn: TESS Peak Height}
\end{align}

\subsection{Case $\psi^{(2)}_\mathrm{a} \approx \psi^{(2)}_\mathrm{b}$}

The situation when the pulses have different GDDs is more complex, but progress can be made by treating $H_\mathrm{a,b}(t,\Omega) $ as a Fourier transform of a product, which is equal to the convolution of the individual Fourier transforms:

\begin{align}
H_\mathrm{a,b}(t,\Omega) &= \frac{1}{\sqrt{2\pi}}\int^{\infty}_{-\infty} \left| \mathcal{E}_\mathrm{a}(\omega - \Omega) \right| \left| \mathcal{E}_\mathrm{b}(\omega ) \right| \exp \left[ \mi \psi_\mathrm{a} (\omega - \Omega) - \mi \psi_\mathrm{b} (\omega)  + \mi \omega t \right] \d \omega \nonumber \\
&= \frac{1}{\sqrt{2\pi}} \int^{\infty}_{-\infty} \left[ \frac{1}{\sqrt{2\pi}}\int^{\infty}_{-\infty} \left| \mathcal{E}_\mathrm{a}(\omega' - \Omega) \right| \left| \mathcal{E}_\mathrm{b}(\omega' ) \right| e^{\mi \omega' t'} \d \omega' \right] \nonumber \\
&~~~~~~~~~~~~~~~~ \cdot \left \lbrace  \frac{1}{\sqrt{2\pi}} \int^{\infty}_{-\infty} \exp \left[ \mi \Delta \psi_\mathrm{a,b} (\omega'',\Omega)  + \mi \omega'' (t-t') \right] \d \omega''\right \rbrace \d t' \nonumber \\
&= \frac{1}{\sqrt{2\pi}} \int^{\infty}_{-\infty}  K_\mathrm{a,b}(t',\Omega) \cdot L_\mathrm{a,b}(t-t',\Omega)~ \d t' \label{Eqn: TESS Peak Convolution}
\end{align}

If we truncate the phase expansion in eqn~\ref{eqn: Difference in spectral phase} at second order, the phase term $\Delta \psi_\mathrm{a,b} (\omega'',\Omega)  + \mi \omega'' (t-t')$ is a real quadratic and so the integral $L_\mathrm{a,b}(t,t',\Omega)$ in Eq~\eqref{Eqn: TESS Peak Convolution} is a Gaussian integral with a purely imaginary argument, or a Fresnel integral, which admits an analytic solution:

\begin{align}
L_\mathrm{a,b}(t-t',\Omega) &= \frac{1}{\sqrt{2\pi}} \exp \left [ + \mi  C(\Omega) + \mi \omega_0 (t - t') \right] \nonumber \\
& ~~~~~~~~~~~~~~~~~~~~~~~~~~~~~ \cdot \int^{\infty}_{-\infty} \exp \left \lbrace \mi A (\omega'' - \omega_0 ) ^2 + \mi \left[ B(\Omega) + t - t' \right] (\omega'' - \omega_0 ) \right \rbrace \d \omega''   \nonumber \\
&= \frac{1}{\sqrt{2\pi}} \sqrt{\frac{\pi \mi}{A}} \exp \left \lbrace -\mi \frac{\left[ B(\Omega) + t - t' \right]^2}{4A} \right \rbrace e^{\mi  C(\Omega)} e^{\mi \omega_0 (t - t')} \nonumber \\
&= \sqrt{\frac{\mi}{2A}} e^{\mi  C(\Omega)} e^{\mi \omega_0 (t - t')} e^{\lambda g(t-t',\Omega)},
\end{align}
where $g(\tau,\Omega) = -\frac{\mi}{4} \left( B(\Omega) + \tau \right)^2 $ and $\lambda = A^{-1} =  2 \left( \psi^{(2)}_\mathrm{a} - \psi^{(2)}_\mathrm{b} \right)^{-1}$. This means that for a given $t$ and $\Omega$ the integral $H_\mathrm{a,b}(t,\Omega)$ is of the form $\int_\Gamma f(x) e^{\lambda g(x)} \d x$ and if the difference between the probe and reference pulse GDDs is sufficiently small, $\lambda \gg 1$. $g(t-t',\Omega)$ is an exact quadratic in $t'$ and so if $K_\mathrm{a,b}(t',\Omega)$ is sufficiently well behaved we can extend $t'$ to the complex plane while ensuring that $f(t',\Omega)$ and $g(t-t',\Omega)$ are holomorphic. This allows us to deform the contour of integration $\Gamma$ and use the method of steepest descent, with $g(t-t',\Omega)$ having a single non-degenerate saddle point at $t-t'_0 = - B(\Omega)$ at which point $g(t - t'_0,\Omega) = 0$ and $g''(t-t'_0,\Omega) = -\frac{\mi}{2}$:

\begin{align}
H_\mathrm{a,b}(t,\Omega) &= \frac{1}{\sqrt{2\pi}}  \sqrt{\frac{\mi}{2A}} e^{\mi C(\Omega) } \int^{\infty}_{-\infty}  K_\mathrm{a,b}(t',\Omega) e^{\mi \omega (t-t') } e^{\lambda g(t-t',\Omega)} \d t' \nonumber \\
&\approx \sqrt{\frac{\mi}{4\pi A}} e^{\mi C(\Omega) } K_\mathrm{a,b}(t'_0,\Omega) e^{\mi \omega_0 (t - t'_0)} \sqrt{\frac{2\pi} {\lambda}} e^{\lambda g(t-t'_0,\Omega)} \left[- g''(t-t'_0,\Omega) \right]^{-\frac{1}{2}} \nonumber \\
&= e^{\mi C(\Omega) } e^{\mi \omega_0 (t - t'_0)} K_\mathrm{a,b}(t'_0,\Omega)  \nonumber \\
&= \frac{1}{\sqrt{2\pi}} e^{\mi C(\Omega) } e^{-\mi \omega_0 B(\Omega)} \int^{\infty}_{-\infty} \left| \mathcal{E}_\mathrm{a}(\omega' - \Omega) \right| \left| \mathcal{E}_\mathrm{b}^*(\omega' ) \right| e^{\mi \omega' \left( t + B(\Omega) \right)} \d \omega' 
\end{align}

This expression is identical to Eqn~\eqref{Eqn: H when A=0} for the case $\psi^{(2)}_\mathrm{a} = \psi^{(2)}_\mathrm{b}$ and hence when the difference between GDDs is small but finite, the integral will again be maximised at approximately $t_\Omega = -B(\Omega) = \psi^{(2)}_\mathrm{a} \Omega + \psi^{(1)}_\mathrm{b} - \psi^{(1)}_\mathrm{a} $ with an amplitude given by Eqn~\eqref{Eqn: TESS Peak Height}.

\subsection{Case $\psi^{(2)}_\mathrm{a} \neq \psi^{(2)}_\mathrm{b}$ for Gaussian $\left| \mathcal{E}_\x(\omega) \right|$} \label{sec: Non-Equal GDDS}

The final tractable case is when pulses a and b can both be assumed to be Gaussian, but not identical, such that they can be described as:

\begin{align}
\left| \mathcal{E}_\x(\omega) \right| &= \mathcal{E}_{\x0} \exp \left[ - \frac{1}{2}\left( \frac{\omega - \omega_0}{\delta \omega_\x} \right)^2 - \sigma_\x (\omega-\omega_0) \right],
\end{align}
where $\delta \omega_\x$ is a measure of the spectral bandwidth of the pulse and $\sigma_\x$ allows the central frequency of the pulse to vary.

Then,
\begin{align}
\left| \mathcal{E}_\a(\omega - \Omega) \right| \left| \mathcal{E}_\b(\omega) \right| &= \mathcal{E}_{\a0} \mathcal{E}_{\b0} \exp \left[ - \frac{1}{2}\left( \frac{\omega - \omega_0 - \Omega}{\delta \omega_\a} \right)^2  - \sigma_\a (\omega-\omega_0) - \frac{1}{2}\left( \frac{\omega - \omega_0}{\delta \omega_\b} \right)^2  - \sigma_\b (\omega-\omega_0)\right] \nonumber \\
&= \mathcal{E}_{\a0} \mathcal{E}_{\b0} \exp \left[ - \frac{1}{2} \left( \frac{1}{\delta \omega_\a^2}  + \frac{1}{\delta \omega_\b^2} \right) \left(\omega - \omega_0 \right)^2  - \left( \sigma_\a + \sigma_\b - \frac{\Omega}{\delta \omega_\a^2 }\right) \left(\omega - \omega_0 \right) - \frac{\Omega^2}{2 \delta \omega_\a^2} \right] \nonumber \\
&= \mathcal{E}_{\a0} \mathcal{E}_{\b0} \exp \left[ - D \left(\omega - \omega_0 \right)^2 - E(\Omega)\left(\omega - \omega_0 \right) - F(\Omega) \right]
\end{align}

This allows us to construct the integrand of $H_\mathrm{a,b}(t,\Omega)$ as a Gaussian, using $\alpha \equiv D -  \mi A$, $\beta(\Omega) \equiv E(\Omega) - \mi B(\Omega) $ and $\gamma \equiv  F(\Omega) - \mi C(\Omega)$, and hence to integrate it exactly, with the positive real part of $\alpha$ ensuring convergence:

\begin{align}
H_\mathrm{a,b}(t,\Omega) &= \frac{1}{\sqrt{2\pi}}\int^{\infty}_{-\infty} \left| \mathcal{E}_\mathrm{a}(\omega - \Omega) \right| \left| \mathcal{E}_\mathrm{b}(\omega ) \right| \exp \left[ \mi \psi_\mathrm{a} (\omega - \Omega) - \mi \psi_\mathrm{b} (\omega)  + \mi \omega t \right] \d \omega \nonumber \\
 &= \frac{1}{\sqrt{2\pi}} \int^{\infty}_{-\infty} \mathcal{E}_{\a0} \mathcal{E}_{\b0} \exp \left[ -\alpha \left(\omega - \omega_0 \right)^2 - \beta(\Omega) \left(\omega - \omega_0 \right) - \gamma (\Omega) + \mi \omega t \right] \d \omega \nonumber \\
&=  \frac{1}{\sqrt{2\pi}} \mathcal{E}_{\a0} \mathcal{E}_{\b0} ~e^{-\gamma (\Omega)} e^{\mi \omega_0 t} \int^{\infty}_{-\infty} \exp \left \lbrace - \alpha \left(\omega - \omega_0 \right)^2 + \left[\mi t - \beta(\Omega) \right] \left(\omega - \omega_0 \right) \right \rbrace \d \omega \nonumber \\
&= \frac{1}{\sqrt{2\pi}} \mathcal{E}_{\a0} \mathcal{E}_{\b0} ~e^{-\gamma (\Omega)} e^{\mi \omega_0 t} \sqrt{\frac{\pi}{\alpha}} \exp \left \lbrace  \frac{\left[ \mi t - \beta(\Omega) \right]^2}{4\alpha}  \right \rbrace \label{Eqn: TESS Peak for Gaussians}
\end{align}

$H_\mathrm{a,b}(t,\Omega)$ is a complex Gaussian and the real part of the argument of the exponential can be written:

\begin{align}
 \Re \left \lbrace \frac{\left[ \mi t - \beta(\Omega) \right]^2}{4\alpha} \right \rbrace &=  \Re \left \lbrace \frac{ \left( D + \mi A\right)}{4\left(A^2 + D^2\right)}  \left[ \mi t - E(\Omega) + \mi B(\Omega) \right]^2 \right \rbrace \nonumber \\
&=  \Re \left( \frac{ \left( D + \mi A\right) }{4\left(A^2 + D^2\right)} \left \lbrace - \left[t  + B(\Omega) \right]^2 + \left[ E(\Omega)\right]^2 - 2 \mi E(\Omega) \left[  t  + B(\Omega) \right] \right \rbrace \right) \nonumber \\
&= \frac{1}{4\left(A^2 + D^2\right)} \left \lbrace - D \left[t  + B(\Omega) \right]^2 + D \left[ E(\Omega)\right]^2 + 2 A E(\Omega) \left[  t  + B(\Omega) \right] \right \rbrace  \nonumber \\
&= \frac{1}{4\left(A^2 + D^2\right)} \left \lbrace - D \left[t  + B(\Omega) - \frac{AE(\Omega)}{D}) \right]^2 +  \frac{\left[A E(\Omega) \right]^2}{D}  + D \left[ E(\Omega)\right]^2 \right \rbrace  \nonumber \\
\end{align}

$H_\mathrm{a,b}(t,\Omega)$ is therefore maximum at,
\begin{align}
t_\Omega &= - B(\Omega) + \frac{AE(\Omega)}{D} \nonumber \\
&= \left( \psi^{(2)}_\a \Omega + \psi^{(1)}_\b - \psi^{(1)}_\mathrm{a} \right) + \frac{1}{2}\left( \psi_\a^{(2)} - \psi_\b^{(2)} \right) \left( \sigma_\a + \sigma_\b - \frac{\Omega}{\delta \omega_\a^2} \right) \left[ \frac{1}{2} \left( \frac{1}{\delta \omega_\a^2} + \frac{1}{\delta \omega_\b^2} \right) \right]^{-1} \nonumber \\
&= \left[ \psi^{(1)}_\b - \psi^{(1)}_\a + \left( \psi_\a^{(2)} - \psi_\b^{(2)} \right) \left( \sigma_\a + \sigma_\b \right) \left( \frac{\delta \omega_\a^2 \delta \omega_\b^2}{\delta \omega_\a^2 + \delta \omega_\b^2} \right)\right] \nonumber \\
& ~~~~ +  \Omega \left[ \psi^{(2)}_\a - \left( \psi_\a^{(2)} - \psi_\b^{(2)} \right)\frac{1}{\delta \omega_\a^2} \left(  \frac{\delta \omega_\a^2 \delta \omega_\b^2}{\delta \omega_\a^2 + \delta \omega_\b^2} \right) \right] \nonumber \\
&= t_0 + \Omega \left[ \left( \delta \omega_\a^2 + \delta \omega_\b^2 \right)  \psi^{(2)}_\a - \delta \omega_\b^2\left( \psi_\a^{(2)} - \psi_\b^{(2)} \right) \right] \left( \frac{1}{\delta \omega_\a^2 + \delta \omega_\b^2}  \right) \nonumber \\
&= t_0 + \Omega \frac{\delta \omega_\a^2 \psi^{(2)}_\a + \delta \omega_\b^2 \psi^{(2)}_\b}{\delta \omega_\a^2 + \delta \omega_\b^2} \\
&= t_0 + \Omega \psi^{(2)}_\mathrm{eff}
\end{align}

This implies that instead of the GDD of pulse a, $\psi^{(2)}_\a$, we should use the mean GDD of pulses a and b, weighted by the square of their bandwidths. If the probe and reference pulses are similar but not identical, $\delta \omega_a \approx \delta \omega_b$, this tends towards the mean  GDD, $\frac{1}{2}\left( \psi^{(2)}_\a + \psi^{(2)}_\b \right)$. While the position of the main sideband $t_0$ is dependent on the central frequencies of the two pulses through $\sigma_a$ and $\sigma_b$, the separation of the peak from the main sideband, $\Omega \psi^{(2)}_\mathrm{eff}$, is unaffected by changes in the central frequency of the two pulses.

\end{document}